\documentclass[english,a4paper,pre,twocolumn,amsmath,amssymb,superscriptaddress]{revtex4-1}
\usepackage{preamble}
\usepackage{xcolor,soul}

\newcommand{\ball}{\mathcal{B}}

\renewcommand{\emptyset}{\varnothing}

\begin{document}

\title{Morphological thermodynamics for hard bodies from a controlled expansion}
\date{\today}

\author{Joshua F. Robinson}
\email{joshua.robinson@bristol.ac.uk}
\affiliation{H.\ H.\ Wills Physics Laboratory, University of Bristol, Bristol BS8 1TL, UK}
\author{Roland Roth}
\affiliation{Institut f\"ur Theoretische Physik, Universit\"at T\"ubingen, 72076 T\"ubingen, Germany}
\author{C. Patrick Royall}
\affiliation{H.\ H.\ Wills Physics Laboratory, University of Bristol, Bristol BS8 1TL, UK}
\affiliation{School of Chemistry, Cantock’s Close, University of Bristol, Bristol, BS8 1TS, UK}
\affiliation{Centre for Nanoscience and Quantum Information, University of Bristol, Bristol BS8 1FD, UK}

\begin{abstract}
  The morphometric approach is a powerful ansatz for decomposing the chemical potential for a complex solute into purely geometrical terms.
  This method has proven accuracy in hard spheres, presenting an alternative to comparatively expensive (classical) density functional theory approaches.
  Despite this, fundamental questions remain over why it is accurate and how one might include higher-order terms to improve accuracy. 
  We derive the morphometric approach as the exact resummation of terms in the virial series, providing further justification of the approach.
  The resulting theory is less accurate than previous morphometric theories, but provides fundamental insights into the inclusion of higher-order terms and to extensions to mixtures of convex bodies of arbitrary shape.
\end{abstract}


\maketitle

\section{Introduction}

The standard theoretical framework for treating inhomogeneous liquids is classical density functional theory (DFT).
Central to this theory is the result that the free energy can be \emph{exactly} expressed as a functional of the density $\Omega = \Omega[\rho(\vec{r})]$ \cite{EvansAP1979}, though approximate functionals must be used in general.
For example, fundamental measure theory (FMT) \cite{RosenfeldPRL1989} provides a class of highly accurate functionals for the hard sphere liquid.
A common practical application of DFT is to its \emph{dual} problem: determining the free energy $\Omega = \Omega[\phi_\mathrm{ext}(\vec{r})]$ for a fixed external potential $\phi_\mathrm{ext}(\vec{r})$.
Approaching this through DFT requires minimisation of $\Omega$ to obtain the equilibrium density profile, a tractable but expensive procedure.
In situations where many function evaluations are required, e.g.\ when integrating over many different realisations of $\phi_\mathrm{ext}$, this minimisation operation can become prohibitively expensive.
It is worthwhile to investigate more direct routes to approximating $\Omega[\phi_\mathrm{ext}(\vec{r})]$, especially where accuracy may be less important than fast calculation.

A promising approach to the dual problem is through morphological thermodynamics \cite{KonigPRL2004} with the potential to enable fast and accurate calculations in hard spheres \cite{RothPRL2006,Hansen-GoosPRL2007,RobinsonPRL2019,RobinsonPRE2019}.
The morphometric approach concerns sharply repulsive external potentials where $\phi_\mathrm{ext}$ acts as a container for the fluid or as an exclusion volume for e.g.\ a solute.
In this limit, the density profile is negligible over a volume $V$ and the free energy is expanded in terms involving $V$ and its boundary $\partial V$.
In this approximation the free energy change from its homogeneous value $\Delta \Omega := \Omega[\phi_\mathrm{ext}] - \Omega_\mathrm{hom}$ is expanded as
\begin{equation}\label{eq:morphometric-approach}
  \Delta \Omega
  =
  p V + a_2 A + a_1 C + a_0 X,
\end{equation}
where $p$ is the pressure and $\{a_2,a_1,a_0\}$ are coefficients for the surface terms $A$, the area; $C$, the integrated mean curvature; and $X$, the integrated Gaussian curvature.
The surface terms are normally determined from the integrals
\begin{subequations}
  \begin{align}
    A &= \int_{\partial V} \, dA \\
    C &= \frac{1}{2} \int \Tr{\kappa} \, dA \\
    X &= \int_{\partial K} \det{\kappa} \, dA
  \end{align}
\end{subequations}
where $\kappa$ is the curvature tensor on the surface.

Despite its accuracy in hard spheres, the morphometric expansion \eqref{eq:morphometric-approach} is still an approximation as been demonstrated in numerous detailed investigations \cite{OettelEL2009,AshtonPRE2011,LairdPRE2012,BlokhuisPRE2013,UrrutiaPRE2014,Hansen-GoosJCP2014,ReindlPRE2015}.
Fundamental questions remain over \emph{why} it is accurate and how one might improve the approximation.
Inaccuracies become significant in hard spheres at very high densities approaching the glass transition \cite{RobinsonPRL2019,RobinsonPRE2019}, so an approximation scheme including additional terms could be desirable for approaches to dynamical arrest.
In this work we will attempt to start the path towards supplementing the morphometric approach \eqref{eq:morphometric-approach} with higher-order terms, by deriving the known terms as the leading contribution in the only properly rigorous free energy expansion: the \emph{virial series}.
This route suggests a properly controlled way of including successive corrections to the approach.
The virial series is dimension-independent so this approach could potentially connect with calculations in high dimensions \cite{LeithallPRE2011,ParisiRMP2010}; though we work in physical dimensions $d \le 3$ it would be straightforward to extend to $d > 3$.

Traditionally, expansions of $\Delta \Omega$ have been obtained in an \emph{ad hoc} way rather than as part of a controlled expansion.
To illustrate this we consider what happens if one attempts to extend \eqref{eq:morphometric-approach} by including higher moments of curvature.
A prototypical example of this is the \emph{Helfrich expansion} for elastic membranes \cite{HelfrichZFNC1973}, which is often argued to be the most general expansion for the surface tension e.g.\ in Ref.\ \cite{BlokhuisPRE2013}.
In this expansion, the next leading order correction to \eqref{eq:morphometric-approach} would be
\begin{equation*}
  \int_{\partial V} \left(\Tr{\kappa}\right)^2 dA,
\end{equation*}
which is not well-defined for general surfaces, in particular for surfaces containing vertices and/or arcs as occurs in e.g.\ polyhedra.
To demonstrate this we consider the line where two planes intersect with dihedral angle $\Delta \theta$.
This can be considered as a cylindrical sector in the limit of vanishing radius $r$, giving the contribution per unit length
\begin{equation*}
  \int \left(\Tr{\kappa}\right)^2 r d\theta
  = \frac{\Delta \theta}{r}
\end{equation*}
diverging as $1 / r$ in the limit where the sector becomes an arc $r \to 0$.
By contrast, the geometric terms already present in the morphometric approach remain finite even where a curvature tensor is not locally definable, at e.g.\ a cusp.
We find that only the curvature terms already present in the morphometric approach are well-defined in general.
Thus, the coefficients of any higher-order moments of curvature must necessarily be zero within a controlled expansion.
The inclusion of higher-order curvatures was originally motivated by continuum elasticity \cite{HelfrichZFNC1973}, so it is not surprising that features on small lengthscales are pathological.

More generally, we find that \emph{any} analytic geometric expansion of $\Delta \Omega$ cannot be exact.
It was shown in the original papers on scaled particle theory \cite{ReissJCP1959,ReissJCP1960} that $\Delta \Omega$ contains singularities, which cannot be captured by simple geometric expansions.
The virial series is in principle exact, so any singularities should be captured by resumming its terms which could suggest new forms for better approximation schemes.

In sections \ref{sec:low-densities} and \ref{sec:hard-rods} we will present the limiting cases where the insertion cost rigorously takes the morphometric form, i.e.\ the low density limit and for one-dimensional hard rods.
In section \ref{sec:finite-densities} we resum the terms contributing in these exact limits to obtain a piece of the solvation energy which \emph{exactly} obeys the morphometric form, and we are able to calculate the thermodynamic coefficients explicitly.
Our main result is valid for hard interactions where the solute and solvent particles are compact and convex.
Though applicable to arbitrary mixtures of particle geometries, the resulting form is equivalent to the standard morphometric approach \eqref{eq:morphometric-approach}.
The methods we use are identical to those used in analysis of inhomogeneous FMT, reflecting the deep underlying connections between FMT and the morphometric approach \cite{LeithallPRE2011,KordenPRE2012,MarechalPRE2014}.
Finally, in section \ref{sec:explicit-lambda} we determine the parameters entering the theory explicitly for hard spheres in physical dimensions $d \le 3$, then in section \ref{sec:numerics} we show (numerically) that the accuracy of this theory is competitive to more traditional hard sphere approaches.

\section{Notation and selected facts from integral geometry}
\label{sec:morph-overview}

We will focus on liquids composed of hard bodies which are \emph{convex} and \emph{compact}.
By compact, we mean objects which are
\begin{enumerate}
\item \emph{bounded}, so they must be finite in scope, as no meaningful size can be defined for a body spanning an infinite region of space, and
\item \emph{closed}, so they contain their boundary.
\end{enumerate}
We write the collection of objects in $d$-dimensions which are compact and convex as $\mathcal{K}^d$.

To obtain results for all physical dimensions $d \le 3$ it is convenient to generalise the morphometric \emph{ansatz} \eqref{eq:morphometric-approach} to arbitrary $d$ and substitute $d \in \{1, 2, 3\}$ at the end of our derivation.
To that end it is convenient to introduce generalisations of the geometric parameters $\{V,A,C,X\}$: the \emph{intrinsic volumes} $\{V_d, V_{d-1}, \cdots, V_0\}$.
To introduce the intuition behind these generalised volumes we start from the observation that the quantities $\{V,A,C,X\}$ can be imagined as the size of projections onto $k$-dimensional subspaces in $\mathbb{R}^3$; for a compact and convex body $K \in \mathcal{K}^3$ we have:
\begin{enumerate}
\item $V[K]$ is trivially the volume of the intersection of $K$ with the 3-dimensional subspace i.e.\ all of Euclidean space.
\item $A[K]$ can be thought of as the typical size of two-dimensional images formed by projections onto planes.
\item $C[K]$ is related to the projections onto one-dimensional subspaces i.e.\ lines.
  This curvature measure is normally thought of as a surface property, but this definition suggests an equivalence (up to a different normalisation) with the \emph{mean width} $L[K]$ of the body.
\item $X[K]$ is obtained from projections onto a single point; this surface measure is thus equivalent to the Euler characteristic $\chi[K]$.
  This connection results in the celebrated Gauss-Bonnet theorem of differential geometry, which we would write
  \begin{equation*}
    X[K] = 2\pi \chi[K]
  \end{equation*}
  in our notation.
\end{enumerate}

\begin{table}
  \begin{ruledtabular}
    \begin{tabular}{ccccc}
      $k$ & $\omega_k$ & $V_k[\ball_1]$ & $V_k[\ball_2]$ & $V_k[\ball_3]$ \\
      \hline
      0 & 1 & 1 & 1 & 1 \\
      1 & 2 & 2 & $\pi$ & 4 \\
      2 & $\pi$ & - & $\pi$ & $2\pi$ \\
      3 & $\frac{4\pi}{3}$ & - & - & $\frac{4\pi}{3}$ \\
    \end{tabular}
  \end{ruledtabular}
  \caption{Intrinsic volumes of the $d$-dimensional unit ball $\ball_d$ in physical dimensions $d \le 3$.}
  \label{table:ball-intrinsic-volumes}
\end{table}

Generalising the above intuition to $d$-dimensions, we see that in general we can imagine $d+1$ projections and so expect $d+1$ corresponding volumes.
We define the $k$th intrinsic volume as the average size of the projections onto $k$-dimensional linear subspaces of $\mathbb{R}^d$, i.e.\ \cite{Klain1997,Santalo2004}
\begin{equation}\label{eq:intrinsic-volumes}
  V_k(K)
  =
  C_{k,d-k}
  \int \chi[K \cap E_{d-k}] \, dE_{d-k}
\end{equation}
where the integral is taken over all affine transformations of the plane $E_{d-k}$ in $\mathbb{R}^d$, and flag coefficient is
\begin{equation}\label{eq:flag-coefficients}
  C_{k,d-k}
  :=
  \frac{d!}{k! (d-k)!} \frac{\omega_d}{\omega_k \omega_{d-k}},
\end{equation}
where the volume of the $d$-dimensional ball with unit radius $\ball_d$ is
\begin{equation}
  \omega_d := V_d[\ball_d] = \frac{\pi^{d/2}}{\Gamma(\frac{d}{2} + 1)}.
\end{equation}
The flag coefficients $C_{k,d-k}$ have a similar structure to binomial coefficients, and play a similar \emph{combinatorial} role in the combination of geometric objects (see kinematic formulae below).
By convention, the normalisation of the measure $dE_{d-k}$ in \eqref{eq:intrinsic-volumes} is chosen to give the intrinsic volumes for the unit ball as
\begin{equation}\label{eq:intrinsic-volume-ball}
  V_k [\ball_d]
  =
  {d \choose k} \frac{\omega_d}{\omega_{d-k}}.
\end{equation}
with values in physical dimensions $d \le 3$ given in Table \ref{table:ball-intrinsic-volumes}.
A set of common geometrical quantities and their reduction to the intrinsic volumes in $d \le 3$ is given in Table \ref{table:geometric-quantities}.
In terms of the intrinsic volumes, the generalisation of the morphometric apparoach \eqref{eq:morphometric-approach} for a solute $K \in \mathcal{K}^d$ in $d$-dimensions reads
\begin{equation}\label{eq:morphometric-approach-d}
  \Delta \Omega[K]
  =
  \sum_{k=0}^d a_k V_k[K].
\end{equation}

\begin{table}
  \begin{ruledtabular}
  \begin{tabular}{ccc}
    Property & Symbol & Functional \\
    \hline
    \multicolumn{3}{c}{$d = 1$} \\
    \hline
    Euler characteristic & $\chi$ & $V_0$ \\
    Length & $L$ & $V_1$ \\
    \hline
    \multicolumn{3}{c}{$d = 2$} \\
    \hline
    Euler characteristic & $\chi$ & $V_0$ \\
    Perimeter & $L$ & $2 V_1$ \\
    Area & $A$ & $V_2$ \\
    \hline
    \multicolumn{3}{c}{$d = 3$} \\
    \hline
    Euler characteristic & $\chi$ & $V_0$ \\
    Mean width & $L$ & $\frac{1}{2} V_1$ \\
    Surface area & $A$ & $2 V_2$ \\
    Volume & $V$ & $V_3$ \\
    Integrated Gaussian curvature & $X$ & $4 \pi V_0$ \\
    Integrated mean curvature & $C$ & $\pi V_1$ \\
  \end{tabular}
  \end{ruledtabular}
  \caption{Common geometrical properties and their representation in terms of the intrinsic volumes $\{V_k\}$.
    The intrinsic volumes are morphological measures describing the size of a body.
    The common geometric interpretations of $V_k$ for $k < d$ typically involves integrations over the boundary $\partial K$ rather than $K$ itself, leading to the curvature measures $\{C,X\}$ in $d=3$ giving an equivalent description as one involving Euler characteristic and the typical width $\{\chi, L\}$.
    However, the intrinsic volumes are more general as they can be evaluated for shapes where curvatures are not locally defined, e.g. at lines and vertices.}
  \label{table:geometric-quantities}
\end{table}

It is usual for liquid state theories to focus on spherically symmetric potentials, so the terms in the virial series involve integrations over particle positions $\{\vec{r}_1, \cdots, \vec{r}_n\}$.
However, integral geometry more naturally deals with non-spherical objects so we can consider this generalisation for the small cost of additional notation.
In addition to translational integrations we also have to consider particle orientations $\{\vec{\theta}_1, \cdots, \vec{\theta}_n\}$ where each $\vec{\theta}_i$ represents an Euler angle tuple.
Then, assuming an isotropic phase where all orientations are equally likely each positional integral generalises to
\begin{equation*}
  \int_{\mathbb{R}^d} d\vec{r}
  \to
  \int_{\mathbb{R}^d \times SO(d)} d\vec{r} d\vec{\theta}
  :=
  \int_{G_d} dg,
\end{equation*}
with the normalisation in the angular measure such that $\int d\vec{\theta} = 1$. In the right-most equality we introduced the rigid motion operation acting on a body $A \subset \mathbb{R}^d$ as
\begin{equation*}
  g A := \{\mathcal{R}(\vec{\theta}) \vec{a} + \vec{r} \, | \, \vec{a} \in A\},
\end{equation*}
a member of the rigid motion group $g \in G_d := \mathbb{R}^d \times SO(d)$, and where $\mathcal{R} \in SO(d)$ is the rotation matrix parameterised by $\vec{\theta}$.
We can take standard results for simple liquids interacting via spherically symmetric pair potentials, and make the above replacement to obtain the correct generalisation for arbitrary shapes.

In evaluating exact contributions in the virial series we will make use of the principal kinematic formula of integral geometry \cite{BlaschkeMZ1936,Blaschke1937,SantaloASI1936}, which express the following \emph{collisional integrals} for $K_1, K_2 \in \mathcal{K}^d$ as \cite{Santalo2004,Klain1997}
\begin{equation}\label{eq:binomial-kinematic-formula}
  \int_{G_d} \chi[K_1 \cap g K_2] \, dg
  =
  \sum_{k=0}^d (C_{k,d-k})^{-1} V_k[K_1] V_{d-k}[K_2].
\end{equation}
The flag coefficients \eqref{eq:flag-coefficients} return here to play an analogous role in conjugating the intrinsic volumes as binomial coefficients do in algebraic expansions.
This formula \eqref{eq:binomial-kinematic-formula} can be iterated for the intersections of many bodies $K_1, \cdots, K_n \in \mathcal{K}^d$ giving \cite{Santalo2004,MarechalPRE2014}
\begin{subequations}\label{eq:multinomial-kinematic-formula}
  \begin{equation}
    \begin{split}
      & \quad
      \int_{G_d^n} \chi[K_1 \cap g_2 K_2 \cap \cdots \cap g_n K_n]
      \, dg_2 \cdots dg_n
      \\ = &
      \sum_{\substack{i_1, \cdots, i_n = 0 \\ i_1 + \cdots + i_n = nd}}^d
      (C_{i_1, \cdots, i_n})^{-1}
      V_{i_1}[K_1]
      \prod_{j=2}^n
      V_{i_j}[K_j],
    \end{split}
  \end{equation}
  \begin{equation}
    \textrm{with} \qquad
    C_{i_1, \cdots, i_n}
    := \frac{1}{i_1! \omega_{i_1}}
    \prod_{j=2}^n
    \left(
    \frac{d!}{i_j!} \frac{\omega_d}{\omega_{i_j}}
    \right),
  \end{equation}
\end{subequations}
where $\int_{G_d^n} dg^n = \int_{G_d} dg_1 \cdots \int_{G_d} dg_n$.
Here $C_{i_1, \cdots, i_n}$ would be the multinomial generalisation of the flag coefficients \eqref{eq:flag-coefficients}.

\section{Exact morphometric limits}

\subsection{One-dimensional hard rods}
\label{sec:hard-rods}

The one-dimensional analogue of a hard sphere is a hard rod
\footnote{In fact, rods are the only convex shape possible in 1d (as line segments) so they are really the one-dimensional analogue of \emph{any} convex object.}.
The cost of inserting a new rod of length $L$ exactly fits the morphometric form independent of density.

Imagine a hard rod fluid occupying all of space.
If we insert a single fixed hard point at the origin, this splits the fluid into two half spaces on either side of the origin, i.e.\ $x < 0$ (left) and $x > 0$ (right).
Because the interactions are hard the two half spaces will be completely decorrelated; thus, growing the point to become a rod of finite size $L$ will simply correspond to translating one of the spaces a distance $L$ requiring work $p L$.
In the limit $L \to 0$ where the rod becomes a point there will be a fixed insertion cost
\begin{equation*}
  \beta \Delta \Omega(L=0) = -\ln{(1-\eta)}
\end{equation*}
coming from the fact that the probability that a randomly chosen position is unoccupied is simply the free volume $1-\eta$ \cite{ReissJCP1959}.
Combining these two terms gives the total cost of inserting a finite sized rod as
\begin{equation}\label{eq:hard-rods-morphometric}
  \beta \Delta \Omega(L) = \beta p L - \ln{(1-\eta)}
\end{equation}
which is exactly of morphometric form (cf.\ \eqref{eq:morphometric-approach-d}).
The pressure can be determined as \cite{TonksPR1936}
\begin{equation}\label{eq:hard-rods-eos}
  \frac{\beta p}{\rho} = \frac{1}{1 - \eta}.
\end{equation}

The morphometric form is violated when multiple rods are inserted at such a distance apart so that a liquid is confined between them.
In this case long-range correlation effects form between the rods which are not captured by the geometric expansion.

\subsection{Low densities in arbitrary dimensions}
\label{sec:low-densities}

We will now obtain the low density asymptotics of the chemical potential, and show that this exactly follows the morphometric form for convex bodies.
This argument follows a line of thought similar to Ref.\ \cite{WertheimMP1994}.

Hard particles feature purely geometric interactions, a property that allows us to make progress.
In particular, the interaction potential between two compact and convex hard bodies $A,B \in \mathcal{K}^d$ is normally written
\begin{equation*}\label{eq:hard-mayer-f}
  u(A,B)
  =
  \begin{cases}
    0 & \textrm{ if } A \cap B = \emptyset \\
    \infty & \textrm{ if } A \cap B \ne \emptyset
  \end{cases}
\end{equation*}
from which the \emph{Mayer function} $f_{AB}$ can be written in the revealing form
\begin{equation*}\label{eq:hard-mayer-f}
  -f_{AB}
  :=
  1 - e^{-\beta u(A,B)}
  =
  \begin{cases}
    0 & \textrm{ if } A \cap B = \emptyset \\
    1 & \textrm{ if } A \cap B \ne \emptyset
  \end{cases}
\end{equation*}
The latter form is identical in form to the Euler characteristic
of their intersection i.e.\
\begin{equation*}
  \chi[A \cap B] =
    \begin{cases}
    0 & \textrm{ if } A \cap B = \emptyset \\
    1 & \textrm{ if } A \cap B \ne \emptyset
    \end{cases}
\end{equation*}
valid for convex bodies.
Comparing this expression with \eqref{eq:hard-mayer-f} we can rewrite the thermodynamic quantity as the purely geometrical measure
\begin{equation}\label{eq:chi-replacement}
  1 - e^{-\beta u(A,B)} = \chi[A \cap B] = -f_{AB}.
\end{equation}
Rewriting the interactions in terms of the Euler characteristic will allow us to exploit the kinematic formulae \eqref{eq:binomial-kinematic-formula} and \eqref{eq:multinomial-kinematic-formula} to evaluate thermodynamic quantities.

Including their relative orientations, the cost of inserting a solute $A$ into a liquid of $B$ particles in the low density limit $\rho \to 0$ is determined from the leading contribution to the virial series using \cite{Hansen2013,Santos2016}
\begin{equation}\label{eq:low-density-insertion}
  \begin{split}
    \beta\Delta\Omega
    &=
    \frac{\rho}{2} \int_{G_d} \left( 1 - e^{-\beta u(A, gB)} \right) dg
    + \mathcal{O}(\rho^2)
    \\ &=
    \frac{\rho}{2} \int_{G_d} \chi[A \cap gB] \, dg
    + \mathcal{O}(\rho^2),
  \end{split}
\end{equation}
where we made the replacement \eqref{eq:chi-replacement} in the second line.
The integrand in the latter line of \eqref{eq:low-density-insertion} can be directly evaluated using the principal kinematic formula \eqref{eq:binomial-kinematic-formula} giving the morphometric form \eqref{eq:morphometric-approach-d} with coefficients
\begin{equation*}
  a_k = \frac{V_{d-k}[B]}{C_{k,d-k}},
\end{equation*}
with coefficients $C_{k,d-k}$ defined in \eqref{eq:flag-coefficients}.
Thus the morphometric approach is exact in the low density limit.
This leads to elegant formulae e.g.\ for $d = 3$ we obtain
\begin{equation*}\label{eq:low-density-morphometric-result}
  \frac{\beta \Delta \Omega}{2 \pi \rho} =
  V[A] X[B] + A[A] C[B] + C[A] A[B] + X[A] V[B]
\end{equation*}
for $\rho \to 0$, where we have used normalisations of intrinsic volumes as surface measures given in Table \ref{table:geometric-quantities}.
The low density result is a classic application of integral geometry to the liquid state, first obtained by Isihara \cite{IsiharaJCP1950}.

\section{Extension to finite densities in arbitrary dimensions}
\label{sec:finite-densities}

We can identify the insertion cost of a solute particle with the chemical potential of a new species of particle (a single solute) in the infinitely dilute limit \cite{ReissJCP1959,Hansen-GoosJPCM2006,Hansen-GoosJCP2014}.
Interestingly, taking this limit for a bulk hard sphere system modelled with fundamental measure theory (FMT) gives the morphometric approach \cite{Hansen-GoosJPCM2006}; this is due to the approximation underlying FMT is that the free energy density can be represented in terms of weighted densities, which are deeply connected to intrinsic volumes.
Alternatively, the exact free energy of this system can be expressed as a virial expansion \cite{Hansen2013}.
This idea was explored in Ref.\ \cite{Hansen-GoosJPCM2006} to show that the morphometric approach \eqref{eq:morphometric-approach-d} is inexact, however here we will attempt a different strategy: we will identify a contribution in the virial expansion which guarantees an insertion cost of morphometric form.
The remaining contributions are unlikely to be rigorously of this form, and their omission is an approximation.

We consider an $(m+1)$-component mixture and we label each species with index $s \in \{0, 1, \cdots, m\}$: the components labelled $\{1, \cdots, m\}$ make up those species present the bulk liquid while the additional component with index $\{0\}$ represents the solute.
Furthermore, we assume each particle in this mixture is a compact and convex body $K_0, K_1, \cdots K_m \in \mathcal{K}^d$.
We will shortly find the chemical potential of the solute by considering the infinitely dilute limit.
The virial series expansion of the excess free energy density is given as \cite{Hansen2013,Santos2016}
\begin{equation}\label{eq:free-energy-density}
  \frac{\beta F^\mathrm{ex}}{V}
  =
  \sum_{n=2}^\infty
  \frac{1}{n-1}
  B_n
  \rho^n,
\end{equation}
with virial coefficients
\begin{equation}
  B_n
  =
  \sum_{s_1=0}^m \cdots \sum_{s_n=0}^m
  B_{s_1, \cdots, s_n} \prod_{i=0}^n x_{s_i},
\end{equation}
where $x_i$ is the mole fraction of species $i$ such that $x_i > 0$ and $\sum_{i=0}^m x_i = 1$.
$B_{s_1, \cdots, s_n}$ are the composition independent virial coefficients describing the contribution from interactions between $n$ particles of species $\{s_0, s_1, \cdots, s_n\}$.
Each contribution contains integrals over all configurations of the $n$ particles \cite{Hansen2013,Santos2016}.
We will refer to these integrals as diagrams because they are normally represented using graph theoretic tools.

We now identify the insertion cost for a new solute particle with its chemical potential in the dilute limit.
The chemical potential of the solute species is
\begin{equation}  
  \beta \mu^\mathrm{ex}_0
  =
  \frac{1}{\rho}
  \frac{\partial}{\partial x_0}
  \left( \frac{\beta F^\mathrm{ex}}{V} \right)_{V,T}
\end{equation}
giving in the dilute limit $x_0 \ll 1$
\begin{equation}\label{eq:chemical-potential-mixture}
  \begin{split}
    \beta \Delta\Omega
    &=
    \lim_{\substack{x_0 \to 0}}
    \sum_{n=2}^\infty
    \frac{1}{n-1}
    \frac{\partial B_n}{\partial x_0}
    \rho^{n-1}
    \\
    &=
    \sum_{n=2}^\infty
    \frac{n}{n-1}
    B_{n-1}^*
    \rho^{n-1}
    =
    \sum_{n=1}^\infty
    \frac{n+1}{n}
    B_n^*
    \rho^n
  \end{split}
\end{equation}
with modified virial coefficient
\begin{equation}
  B_n^* =
  \sum_{s_1=1}^m \cdots \sum_{s_n=1}^m
  B_{0, s_1, \cdots, s_n}
  \prod_{i=1}^n x_{s_i}
\end{equation}
which contains contributions from all diagrams containing a single member of the solute species.

We now introduce our central approximation which generically results in a morphometric form for $\beta \Delta \Omega$, for arbitrary mixtures of hard particles and in all densities and dimensions: we select only contributions to $B_{0, s_1, \cdots, s_n}$ where there is a common point of intersection between the $n+1$ particles.
The intuition behind this approximation can be understood by considering again the two limits where the morphometric approach is rigorously exact.
First, in the low density limit the integral \eqref{eq:binomial-kinematic-formula} selects only those geometries where the solute and solvent particle overlap.
Second, in the one-dimensional limit all of the nonzero contributions to the virial expansion occur where there is a common point of intersection \cite{MarechalPRE2014}.
This approximation scheme has been systematically explored in the more general case of inhomogeneous systems \cite{LeithallPRE2011,KordenPRE2012,MarechalPRE2014}, and can be used to derive FMT from first-principles.
In the context of FMT for a binary mixture, the approximate third virial coefficient produced by this method could be highly inaccurate especially for asymmetric particle sizes \cite{CuestaJPCM2002}.
This approximation allows us to write \eqref{eq:chemical-potential-mixture} as
\begin{equation}\label{eq:insertion-with-lambda}
  \beta \Delta \Omega
  =
  \sum_{n=1}^\infty
  c_n
  \rho^n
  \sum_{s_1=1}^m \cdots \sum_{s_n=1}^m
  \Lambda_{s_1, \cdots, s_n}
  \prod_{i=1}^n x_{s_i}
\end{equation}
where $c_n$ is a combinatorial prefactor independent of the interactions or dimensionality, and
\begin{equation}\label{eq:n-particle-intersection-integral}
  \Lambda_{s_1, \cdots, s_n}
  =
  \int_{G_d^n} dg^n \chi[K_0 \cap g_1 K_{s_1} \cap \cdots \cap g_n K_{s_n}]
\end{equation}
counts the number of microstates where there is a region of mutual overlap.
This expression $\Lambda_{s_1, \cdots, s_n}$ is a real contribution in the $n$-particle diagrams of the full virial expansion, and the only approximation here is in neglecting additional terms; this feature makes the resulting theory part of a controlled approximation.

The iterated kinematic formula \eqref{eq:multinomial-kinematic-formula} gives the explicit value of the intersections of many bodies $\{K_i\}$ as \cite{Santalo2004,MarechalPRE2014}
\begin{subequations}\label{eq:multinomial-kinematic-equation}
  \begin{equation}
    \Lambda_{s_1, \cdots, s_n}
    =
      \sum_{\substack{i_0, \cdots, i_n = 0 \\ i_0 + \cdots + i_n = nd}}^d
      (C_{i_0, \cdots, i_n})^{-1}
      V_{i_0}[K_0]
      \prod_{j=1}^n
      V_{i_j}[K_{s_j}],
  \end{equation}
  \begin{equation}
    \textrm{with} \qquad
    C_{i_0, \cdots, i_n}
    := \frac{1}{i_0! \omega_{i_0}}
    \prod_{j=1}^n
    \left(
    \frac{d!}{i_j!} \frac{\omega_d}{\omega_{i_j}}
    \right).
  \end{equation}
\end{subequations}
Introducing the rescaled volumes
\begin{equation}\label{eq:rescaled-intrinsic-volumes}
  \widetilde{V}_k[K_s]
  =
  \frac{k! \omega_k}{d! \omega_d} V_k[K_s]
\end{equation}
eliminates the combinatorial factor in \eqref{eq:n-particle-intersection-integral} giving
\begin{equation}\label{eq:lambda-reduced}
  \Lambda_{s_1, \cdots, s_n}
  =
  d! \omega_d
  \sum_{\substack{i_0, \cdots, i_n = 0 \\ i_0 + \cdots + i_n = nd}}^d
  \widetilde{V}_{i_0}[K_0]
  \prod_{j=1}^n
  \widetilde{V}_{i_j}[K_{s_j}].
\end{equation}
Summing this equation over all the different species in the mixture gives
\begin{equation}
  \label{eq:final-lambda}
  \begin{split}
    \Lambda^*
    :=&
    \sum_{s_1, \cdots s_n = 1}^m
    \Lambda_{s_1, \cdots, s_n}
    \prod_{i=1}^n x_{s_i}
    \\ =&
    d! \omega_d
    \sum_{\substack{i_0, \cdots, i_n = 0 \\ i_0 + \cdots + i_n = nd}}^d
    \widetilde{V}_{i_0}[K_0]
    \prod_{j=1}^n
    \xi_{i_j}
    \\ =&
    d! \omega_d
    \sum_{k = 0}^d
    \frac{\lambda_k^{(n)}}{\rho^n}
    \widetilde{V}_{k}[K_0],
  \end{split}
\end{equation}
where we simplified the final expression by introducing the \emph{scaled particle variables}
\begin{subequations}
  \begin{align}
    \label{eq:spt-variables-resummation}
    \xi_k
    &=
    \rho \sum_{s = 1}^m x_s \widetilde{V}_k[K_s]
    \\
    \label{eq:little-lambda}
    \lambda_k^{(n)}
    &=
    \sum_{\substack{i_1, \cdots, i_n = 0 \\ i_1 + \cdots + i_n = nd - k}}^d
    \prod_{j=1}^n
    \xi_{i_j}.
  \end{align}
\end{subequations}
The bulk volume fraction is generically $\eta = \xi_d$, and the Euler characteristic of the particles must be unity for convex particles giving $\xi_0 = \rho / (d! \omega_d)$.

At this point we can observe that the resulting free energy is already of morphometric form, as can be seen by combining \eqref{eq:insertion-with-lambda}, \eqref{eq:rescaled-intrinsic-volumes} and \eqref{eq:final-lambda} giving
\begin{subequations}\label{eq:morphometric-approach-from-virial}
  \begin{align}
    \beta \Delta \Omega
    &=
    \sum_{k=0}^d \beta a_k V_k[K_0]
    \\ \textrm{with} \qquad
    \beta a_k
    &=
    k! \omega_k \sum_{n=1}^\infty c_n \lambda_k^{(n)}.
    \label{eq:a-coefficient}
  \end{align}
\end{subequations}
This is our main result: the morphometric approach \eqref{eq:morphometric-approach-d} is a well-founded \emph{ansatz} for the insertion free energy, irrespective of the details of the liquid composition and the shapes of the constituent particles.
Next, we will obtain explicit forms of the thermodynamic coefficients $a_k$ to show that this route captures a significant leading contribution to this free energy, reinforcing the strength of the \emph{ansatz}.
To find explicit expressions for $a_k$ we have to determine the combinatorial prefactor $c_n$ and evaluate the geometric/mixture contribution $\lambda_k^{(n)}$.

To obtain the combinatorial coefficient $c_n$ we use a technique suggested in Ref.\ \cite{MarechalPRE2014}: the coefficients are independent of dimensionality so we can compare the form of \eqref{eq:morphometric-approach-from-virial} against the exact free energy known for $d=0$.
The (quasi--) zero dimensional limit can be thought of as a small cavity which is only able to fit a single particle, as the system size approaches the particle size $V \sim \xi_d$.
The exact free energy is known to be \cite{RosenfeldJPCM1996,MarechalPRE2014}
\begin{equation}\label{eq:free-energy-zero-d-1}
  \begin{split}
    \lim_{d \to 0}
    \beta F^\mathrm{ex}
    &=
    (1 - \rho V) \ln{(1 - \rho V)} + \rho V
    \\ &=
    \sum_{n=2}^\infty \frac{(\rho V)^n}{n(n-1)}
  \end{split}
\end{equation}
where $\rho V < 1$ is the average occupancy of the cavity.
To make comparison with our expression for the chemical potential, we observe that the $k=d$ term in \eqref{eq:morphometric-approach-from-virial} involves the volume of the inserting particle $\widetilde{V}_d(K_0)$ so its conjugate variable must be the pressure $a_d = \beta p$ \cite{ReissJCP1959}.
Explicit evaluation in the $d \to 0$ limit then gives
\begin{equation*}
  \lim_{d \to 0}
  \left(
  \frac{\beta p}{\rho} - 1
  \right)
  =
  \sum_{n=2}^\infty c_n (\rho V)^{n-1}
\end{equation*}
where we recognised $c_1 = 1$ for consistency with the ideal gas law, from which we can obtain the excess free energy through the thermodynamic relation
\begin{equation}\label{eq:free-energy-zero-d-2}
  \begin{split}
    \lim_{d \to 0}
    \beta F^\mathrm{ex}
    &=
    \rho V \int_0^\rho
    \lim_{d \to 0}
    \left(
    \frac{\beta p}{\rho'} - 1
    \right)
    \, \frac{d\rho'}{\rho'}
    \\ &=
    \sum_{n=2}^\infty
    \frac{c_n}{n-1} (\rho V)^n.
  \end{split}
\end{equation}
Comparing \eqref{eq:free-energy-zero-d-1} and \eqref{eq:free-energy-zero-d-2} allows us to read off the combinatorial term as
\begin{equation}\label{eq:c-coefficient}
  c_n = \frac{1}{n}.
\end{equation}
Finally, collecting terms of the same index in \eqref{eq:little-lambda} gives the more tractable sum \cite{MarechalPRE2014}
\begin{equation}\label{eq:horrible-lambda-sum}
  \lambda_k^{(n)}
  =
  \sum_{\substack{
      N_0, N_1, \cdots, N_d \ge 0 \\
      d N_0 + (d-1)N_1 + \cdots + N_{d-1} = k \\
      N_0 + N_1 + \cdots + N_d = n}}^d
  n!
  \prod_{j=0}^d
  \frac{\xi_{j}^{N_j}}{N_j!}
\end{equation}
with the factorials accounting for the different combinations of terms.

Despite the complicated form of the summation limits in \eqref{eq:horrible-lambda-sum}, there are very few contributing terms in physical dimensions $d \le 3$; we will work these out explicitly in the next section.
Resumming the $\lambda_k^{(n)}$ terms gives the thermodynamic coefficients $a_k$ through \eqref{eq:a-coefficient} after inserting the combinatorial term \eqref{eq:c-coefficient} giving
\begin{equation}\label{eq:final-a-coefficient}
  \beta a_k = k! \omega_k \sum_{n=1}^\infty \frac{\lambda_k^{(n)}}{n}.
\end{equation}
Notably, $a_d = p$ gives the equation of state.

\section{Explicit morphometric coefficients from the virial series}
\label{sec:explicit-lambda}

We will now resum $\lambda_k^{(n)}$ over $n$ in order to determine the values of $a_k$ for $d \le 3$.
In the appendix we evaluate the sums in \eqref{eq:horrible-lambda-sum} to obtain $\lambda_k^{(n)}$ for each $d$.

For $d=1$ we insert \eqref{eq:little-lambda-d=1} into \eqref{eq:final-a-coefficient}:
\begin{subequations}
  \begin{align}
    \beta a_0
    &= - \ln{(1 - \xi_1)},
    \\
    \beta p =
    \beta a_1 &=
    \frac{2 \xi_0}{1-\xi_1},
  \end{align}
\end{subequations}
with $\xi_0 = \rho / 2$ and $\xi_1 = \eta$ giving the exact result for hard rods \eqref{eq:hard-rods-morphometric} and \eqref{eq:hard-rods-eos}.

For $d=2$ we insert \eqref{eq:little-lambda-d=2} into \eqref{eq:final-a-coefficient}:
\begin{subequations}
  \begin{align}
    \beta a_0 &= -\ln{(1 - \xi_2)},
    \\
    \beta a_1 &= \frac{2 \xi_1}{1-\xi_2},
    \\
    \beta p =
    \beta a_2 &=
    2\pi \left(
    \frac{\xi_0}{1 - \xi_2}
    + \frac{\xi_1^2}{2(1-\xi_2)^2}
    \right).
  \end{align}
\end{subequations}
For hard discs of diameter $\sigma$ we obtain $V_1 = \pi \sigma / 2$ so $\xi_0 = \rho / (2\pi)$, $\xi_1 = \rho \sigma / 2$ and $\xi_2 = \eta$ for the single-component fluid, which produces coefficients identical to two-dimensional scaled particle theory.
In this limit the chemical potential of inserting a disc of radius $R$ becomes
\begin{equation*}
  \Delta\Omega(R)
  =
  \frac{1}{\sigma^2 (1-\eta)^2} \, \pi R^2
  + \frac{4 \eta}{\sigma (1 - \eta)} R
  - \ln{(1 - \eta)}.
\end{equation*}
It is straightforward to verify by insertion that this satisfies the two-dimensional equivalents of the scaled particle conditions \cite{ReissJCP1959}, demonstrating that this is indeed the scaled particle solution.

Finally, for $d=3$ we insert \eqref{eq:little-lambda-d=3} into \eqref{eq:final-a-coefficient}:
\begin{subequations}\label{eq:ak-d=3}
  \begin{align}
    \beta a_0 &= -\ln{(1 - \xi_3)},
    \\
    \beta a_1 &= \frac{2 \xi_2}{1-\xi_3},
    \\
    \beta a_2 &=
    2 \pi
    \left(
    \frac{\xi_1}{1-\xi_3}
    + \frac{\xi_2^2}{2(1-\xi_3)^2}
    \right),
    \\
    \beta p =
    \beta a_3 &=
    8 \pi
    \left(
    \frac{\xi_0}{1-\xi_3}
    + \frac{\xi_1 \xi_2}{(1-\xi_3)^2}
    + \frac{\xi_2^3}{3 (1-\xi_3)^3}
    \right).
    \label{eq:a3-d=3}
  \end{align}
\end{subequations}
For hard spheres of diameter $\sigma$ we obtain $V_1 = 2\sigma$ and $V_2 = \pi \sigma^2 / 2$ so $\xi_0 = \rho / (8\pi)$, $\xi_1 = \rho \sigma / (2\pi)$, $\xi_2 = \rho \pi \sigma^2 / 8$ and $\xi_3 = \eta$ for the single-component fluid.
In contrast to the $d=2$ case, this result differs from the scaled particle solution for hard spheres.
In $d=3$ the scaled particle theory solution is equivalent to the Percus-Yevick (PY) equation of state obtained by the compressibility route \cite{WertheimPRL1963,LebowitzPR1964}; with our normalisations this solution reads \cite{Santos2016}
\begin{equation}\label{eq:py-pressure-mixtures}
  \beta p^\mathrm{PY} =
  8 \pi
  \left(
  \frac{\xi_0}{1-\xi_3}
  + \frac{\xi_1 \xi_2}{(1-\xi_3)^2}
  + \frac{16 \xi_2^3}{3 \pi^2 (1-\xi_3)^3}
  \right),
\end{equation}
which is exact up to the third virial coefficient $B_3$.
The third term in \eqref{eq:a3-d=3} differs from this by a factor $\pi^2/16$, so the resummation approach only provides a lower bound on $B_3$; the omitted configurations are known in the FMT literature as ``lost cases'' \cite{TarazonaPRE1997}.
However, \eqref{eq:py-pressure-mixtures} applies specifically for hard sphere mixtures whereas the resummation approach \eqref{eq:ak-d=3} is valid for convex/compact particles of arbitrary shape.

\section{Numerical results for single-component hard spheres}
\label{sec:numerics}

\begin{figure}
  \includegraphics[width=\linewidth]{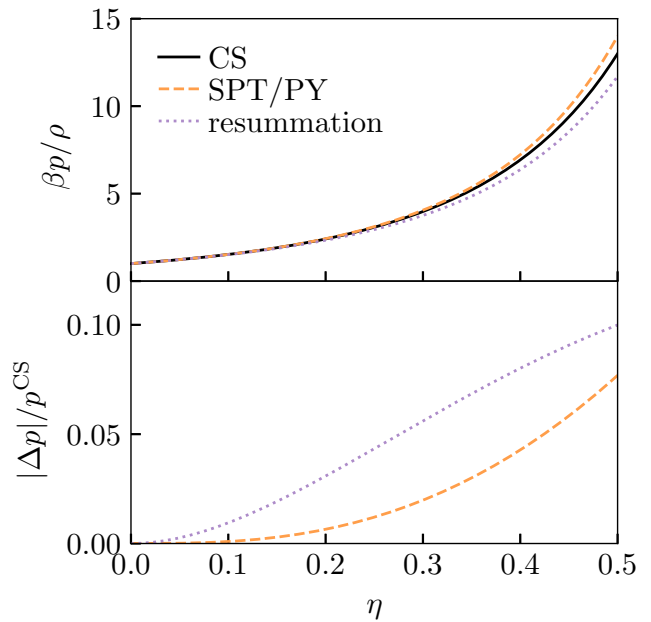}
  \caption{(colour online) Equations of state for the single-component hard sphere liquid in $d=3$: Carnahan-Starling (CS), scaled particle/Percus-Yevick (SPT/PY) and the equation obtained from resumming terms in the virial series where there is a common point of intersection.
    Top panel: pressure equations of state.
    Bottom panel: errors in the SPT/PY and resummation pressures are comparable across the whole liquid regime, taking the CS equation as the quasi-exact result.}
  \label{fig:resummation-pressure}
\end{figure}

For single-component hard spheres the pressure obtained from the resummation in the previous section yields
\begin{equation}
  \frac{\beta p}{\rho} =
  \begin{cases}
    \frac{1}{1-\eta} & \; d=1 \\
    \frac{1}{(1-\eta)^2} & \; d=2 \\
    \frac{1 + \eta + (\frac{3\pi^2}{16} - 2) \eta^2}{(1-\eta)^3} & \; d=3.
  \end{cases}
\end{equation}
The resulting pressures for $d \le 2$ are identical to the scaled particle solutions, and the first is exact \eqref{eq:hard-rods-eos}.
For $d=3$ the resulting equation of state has a similar structure to the scaled particle solution but it is slightly less accurate: at the freezing point $\eta_f \simeq 0.494$ the scaled particle theory solution (SPT/PY) overestimates the pressure by $\sim7\%$ while for the above equation this is underestimated by $\sim11\%$, taking the Carnahan-Starling (CS) equation of state \cite{CarnahanJCP1969} as an estimate of the exact value.

The three equations of state mentioned above for $d=3$ are plotted together in Fig.\ \ref{fig:resummation-pressure} across the whole liquid regime in hard spheres.
While not exact, this shows that the morphometric contributions account for $\sim$90\% of the contributions to the equation of state.
This fact suggests that the reported accuracy of morphological thermodynamics for descriptions of the hard sphere liquid \cite{RothPRL2006,LairdPRE2012,BlokhuisPRE2013,UrrutiaPRE2014,Hansen-GoosJCP2014,RobinsonPRL2019} is possible because this exact contribution is a significant leading contribution.
This is discussed in more detail in the context of FMT in \cite{MarechalPRE2014}, and is partially attributable to cancellations of terms omitted from the resummation.

\begin{figure}[b]
  \includegraphics[width=\linewidth]{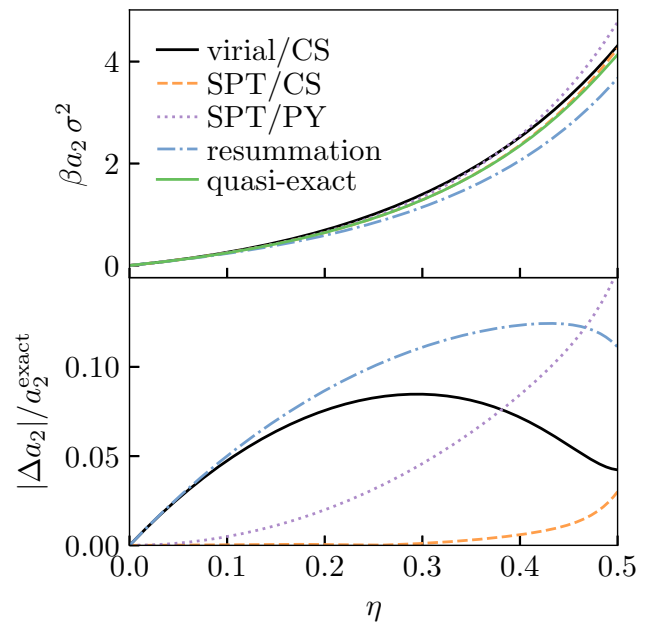}
  \caption{(colour online) Comparison of surface tensions for different morphometric theories.
    using the highly accurate result \eqref{eq:quasi-exact-surface-tension} from Ref.\ \cite{DavidchackMP2015} valid until $\eta \sim 0.5$.}
  \label{fig:resummation-a2}
\end{figure}

We see similar accuracy in the predicted surface tension at an infinite planar wall determined by $a_2$
\footnote{This is true up to a normalisation constant, as $a_2$ conjugates with the intrinsic volume $V_2$ rather than the area $A = 2V_2$.
  The usual planar surface tension is thus obtained as $\gamma_\infty = a_2/2$.}.
To measure accuracy we restate the quasi-exact result \eqref{eq:quasi-exact-surface-tension} of Ref.\ \cite{DavidchackMP2015} in terms of the normalisations used in this work as
\begin{widetext}
\begin{equation}\label{eq:quasi-exact-surface-tension}
  \beta a_2
  =
  \frac{2}{\pi \sigma^2} \left(
  \frac{\eta (2 + 3\eta - \frac{9}{5}\eta^2 - \frac{4}{5}\eta^3 - (5 \times 10^4) \eta^{20})}{(1 - \eta)^2}
  - \ln{(1 - \eta)}
  \right).
\end{equation}
\end{widetext}
We also compare the values of $a_2$ predicted with other morphometric theories which impose the more accurate Carnahan-Starling (CS) equation of state from Refs.\ \cite{Hansen-GoosJPCM2006,RobinsonPRE2019} obtained from scaled particle theory (SPT/CS) and the virial theorem (virial/CS).
In Fig.\ \ref{fig:resummation-a2} we see the surface tensions for each theory across the liquid regime; the accuracy of the new result is comparable to SPT/PY in the liquid regime with the maximum error reaching $\sim12\%$.
Unsurprisingly, the other morphometric theories feature more accurate surface tensions; this is likely because they were specifically constructed to satisfy thermodynamic relations which improves their self-consistency.
Curiously, the error in the new theory scales almost identically to virial/CS theory at small $\eta$ even though it has the opposing sign; all of the previous morphometric theories overestimate the surface tension, whereas the resummation route underestimates it.

\section{Conclusions}

We have derived an exact morphometric contribution for a general class of hard particle liquids by resumming terms in the virial series.
Previous studies have primarily used fundamental measure theory to develop morphometric theories, so we have successfully developed an independent justification for the morphometric approach as the leading term in a controlled expansion.
The exact result applies for mixtures of hard convex particles in an isotropic phase.

In hard spheres, this exact contribution features similar accuracy as scaled particle theory, and exactly coincides with it for $d \le 2$.
Numerical comparison in $d=3$ shows that the pressure and surface tension are comparable in accuracy to the classic scaled particle route, so it captures the dominant contributions to the bulk free energy across a large density range; this latter fact seems to suggest why the approach has been successful.
Though as noted in Ref.\ \cite{MarechalPRE2014}, this is partially due to a cancellation in the omitted terms of the virial expansion.
Finally, we note that our explicit calculations were performed in physical dimensions $d \le 3$, yet our formalism is dimension independent; it would be straightforward to extend the summations to higher dimensions and connect with previous theories e.g.\ Ref.\ \cite{LeithallPRE2011}.

The free energy we have identified emerges \emph{rigorously} as a contribution from the virial series, and its accuracy indicates that the success of morphological theories reported in previous investigations \cite{RothPRL2006,LairdPRE2012,BlokhuisPRE2013,UrrutiaPRE2014,Hansen-GoosJCP2014,RobinsonPRL2019} is enabled by this being a significant leading contribution.
Moreover, the exact contribution provides a suitable starting point for including additional terms where improved accuracy is needed at e.g.\ high densities approaching dynamical arrest.
We could write the insertion cost for a solute $K$ as the \emph{exact} decomposition
\begin{equation}\label{eq:exact-morph-decomposition}
  \Delta \Omega[K]
  =
  \sum_{k=0}^d a_k V_k[K]
  + \Delta \Omega_\mathrm{extra}[K]
\end{equation}
with coefficients $a_k$ as previously calculated, and $\Delta \Omega_\mathrm{extra}$ containing the subleading corrections.
Notably, the exponentially damped oscillations occurring in pair correlations at asymptotically large separations must be contained within $\Delta \Omega_\mathrm{extra}$.
The insertion cost is known to contain singularities \cite{ReissJCP1959} so it is unlikely that $\Delta \Omega_\mathrm{extra}$ possesses a simple analytic form.
It is possible that additional exact morphometric contributions exist, and they would be contained in $\Delta \Omega_\mathrm{extra}$ also.
Furthermore, the formal derivation we have followed naturally leads to explicit expressions for $\Delta \Omega_\mathrm{extra}$.

The next leading contribution from the virial series would be:
\begin{equation*}
  \begin{split}
    \Delta \Omega_\mathrm{extra}[K]
    =&
    \frac{\rho^2}{2}
    \sum_{s_1=1}^m \sum_{s_2=1}^m
    x_{s_1} x_{s_2}
    \left(
    \Delta_{s_1,s_2}
    - \Lambda_{s_1,s_2} \right)
    + \mathcal{O}(\rho^3),
  \end{split}
\end{equation*}
where $\Delta_{s_1,s_2}$ is the three-body \emph{ring integral}.
Ring integrals can be calculated straightforwardly in hard spheres \cite{MontrollJCP1941}, or using the Radon transform for convex geometries of arbitrary shapes \cite{WertheimMP1994,WertheimMP1996,WertheimMP1996a}.
Corrections to the morphometric approximation could be systematically included by further resummations over other classes of diagrams, with ring integrals as the leading order terms.
These corrections are discussed in Ref.\ \cite{MarechalPRE2014} in the context of free energy functionals for inhomogeneous liquids; our system is effectively homogeneous so we expect it to be easier to construct a theory containing these higher-order terms.
Notably, the ring integrals are argued to be the sole contributions in the mean-field infinite-dimensional limit \cite{ParisiRMP2010}.
Resumming the ring diagrams would lead to a contribution in $\Delta \Omega$ involving a double volume integral over the solute geometry, and their inclusion could connect the morphometric approach with the mean-field theory of hard spheres.

The form of the exact contribution is instructive in how it applies to mixtures.
It is argued in Ref.\ \cite{KodamaJCP2011} that for an $m$-component mixture the appropriate morphometric form reads
\begin{equation*}\label{eq:morphometric-approach-mixtures}
  \Delta \Omega[K]
  =
  \sum_{i=1}^m
  a_3^{(i)} V_i[K]
  + a_2^{(i)} A_i[K]
  + a_1^{(i)} C_i[K]
  + a_0^{(i)} X_i[K]
\end{equation*}
where the coefficients $a_k^{(i)}$ now depend on the specific interactions with each species and their composition, and $\{V_i, A_i, C_i, X_i\}$ are geometric measures on some composite body of the solute with solvent particles of species $i$, e.g.\ their specific excluded volume.
By contrast, our exact morphometric contribution does not involve different intrinsic volumes for the different cross-species interactions, suggesting the normal morphometric approach \eqref{eq:morphometric-approach-d} is a general enough \emph{ansatz} and the extension for mixtures proposed in Ref.\ \cite{KodamaJCP2011} may be unnecessary.
Moreover, as functions of the scaled particle variables $\{\xi_i\}$ the coefficients we derive remain well-defined in the polydisperse limit $m \to \infty$, detailed discussion of which can be found in Refs.\ \cite{GualtieriJCP1982,WarrenPRL1998,SollichPRL1998,SollichAiCP2001}.
With the number of coefficients growing with $m$ in the alternative \emph{ansatz} above, it is unclear how well-posed it would be in that limit.

\appendix*

\section{Evaluating sums in the virial series}
\label{appendix:lambda}

Here we explicitly evaluate the contributions in the virial expansion from configurations sharing a common point of intersection via \eqref{eq:horrible-lambda-sum}.

For $d=1$ the index runs over $k \in \{0,1\}$, so the criteria on the summation indices is that $N_0 = k$ and $N_1 = n - N_0$ leading to a single term for each value of $k$:
\begin{subequations}
  \label{eq:little-lambda-d=1}
  \begin{align}
    \lambda_0^{(n)} &= \xi_1^n,
    \\
    \lambda_1^{(n)} &= n \xi_0 \xi_1^{n-1}.
  \end{align}
\end{subequations}
For $d=2$ we have $k \in \{0,1,2\}$, with summation conditions $2N_0 + N_1 = k$ and $N_2 = n - N_1 - N_0$ giving:
\begin{subequations}
  \label{eq:little-lambda-d=2}
  \begin{align}
    \lambda_0^{(n)} &= \xi_2^n,
    \\
    \lambda_1^{(n)} &= n \xi_1 \xi_2^{n-1},
    \\
    \lambda_2^{(n)} &=
    n \xi_0 \xi_2^{n-1}
    + \frac{n(n-1)}{2} \xi_1^2 \xi_2^{n-2}.
  \end{align}
\end{subequations}
Finally, for $d=3$ we have $k \in \{0,1,2,3\}$, with summation conditions $3N_0 + 2N_1 + N_2 = k$ and $N_3 = n - N_2 - N_1 - N_0$ giving:
\begin{subequations}
  \label{eq:little-lambda-d=3}
  \begin{align}
    \lambda_0^{(n)} &= \xi_3^n,
    \\
    \lambda_1^{(n)} &= n \xi_2 \xi_3^{n-1},
    \\
    \lambda_2^{(n)} &=
    n \xi_1 \xi_3^{n-1}
    + \frac{n(n-1)}{2} \xi_2^2 \xi_3^{n-2},
    \\
    \lambda_3^{(n)} &=
    n \xi_0 \xi_3^{n-1}
    + n(n-1) \xi_1 \xi_2 \xi_3^{n-2}
    \nonumber \\ & \qquad
    + \frac{n(n-1)(n-2)}{6} \xi_2^3 \xi_3^{n-3}.
  \end{align}
\end{subequations}

\begin{acknowledgements}
  JFR and CPR  acknowledge the European Research Council under the FP7 / ERC Grant Agreement No.\ 617266 ``NANOPRS''. CPR would like to acknowledge the Royal Society for financial support.
\end{acknowledgements}

\bibliography{bibliography}

\end{document}